\documentclass[10pt, prd, twocolumn, nofootinbib,preprint,superscriptaddress]{revtex4}
\pdfoutput=1
\usepackage[T1]{fontenc}
\usepackage{amsmath,amssymb}
\usepackage{epsfig}
\usepackage{graphicx}
\usepackage[usenames,dvipsnames]{color}
\usepackage{subfigure}
\usepackage{slashed}
\usepackage[colorlinks,citecolor=blue]{hyperref}
\usepackage{pdfpages}
\usepackage{color}

\begin{document}
\title{Inelastic Fermion Dark Matter Origin of XENON1T Excess with \\ Muon $(g-2)$ and Light Neutrino Mass}

\author{Debasish Borah}
\email{dborah@iitg.ac.in}
\affiliation{Department of Physics, Indian Institute of Technology Guwahati, Assam 781039, India}

\author{Satyabrata Mahapatra}
\email{ph18resch11001@iith.ac.in}
\affiliation{Department of Physics, Indian Institute of Technology Hyderabad, Kandi, Sangareddy 502285, Telangana, India}

\author{Dibyendu Nanda}
\email{dibyendu.nanda@iitg.ac.in}
\affiliation{Department of Physics, Indian Institute of Technology Guwahati, Assam 781039, India}

\author{Narendra Sahu}
\email{nsahu@iith.ac.in}
\affiliation{Department of Physics, Indian Institute of Technology Hyderabad, Kandi, Sangareddy 502285, Telangana, India}

\begin{abstract}
Motivated by the recently reported excess in electron recoil events by the XENON1T collaboration, we propose an inelastic fermion dark matter (DM) scenario within the framework of a gauged $L_{\mu}-L_{\tau}$ extension of the standard model which can also accommodate tiny neutrino masses as well as anomalous muon magnetic moment $(g-2)_{\mu}$. A Dirac fermion DM, naturally stabilised due to its chosen gauge charge, is split into two pseudo-Dirac mass eigenstates due to Majorana mass term induced by singlet scalar which also takes part in generating right handed neutrino masses responsible for type I seesaw origin of light neutrino masses. The inelastic down scattering of heavier DM component can give rise to the XENON1T excess for keV scale mass splitting with lighter DM component. We fit our model with XENON1T data and also find the final parameter space by using bounds from $(g-2)_{\mu}$, DM relic, lifetime of heavier DM, inelastic DM-electron scattering rate, neutrino trident production rate as well as other flavour physics, astrophysical and cosmological observations. A tiny parameter space consistent with all these bounds and requirements will face further scrutiny in near future experiments operating at different frontiers.
\end{abstract}

\maketitle

\noindent
{\bf Introduction}: XENON1T collaboration has recently reported an excess of electron recoil events near 1-3 keV energy \cite{Aprile:2020tmw}. While this excess is consistent with the solar axion model at $3.5\sigma$ significance and with neutrino magnetic moment signal at $3.2\sigma$ significance, both these interpretations are in strong tension with stellar cooling constraints. While XENON1T collaboration can neither confirm or rule out the possible origin of this excess arising due to beta decay occurring in trace amount of tritium present in the xenon container, it has generated a great deal of interest among the particle physics community to look for possible new physics interpretations. Different dark matter (DM) interpretations of this excess have been proposed in several works \cite{Smirnov:2020zwf, Takahashi:2020bpq, Alonso-Alvarez:2020cdv, Kannike:2020agf, Fornal:2020npv, Du:2020ybt, Su:2020zny, Harigaya:2020ckz, Chen:2020gcl, Bell:2020bes, Dey:2020sai, Cao:2020bwd, Lee:2020wmh, Paz:2020pbc, Choi:2020udy, An:2020bxd, Baryakhtar:2020rwy, Bramante:2020zos, Jho:2020sku, Nakayama:2020ikz, Primulando:2020rdk, Zu:2020idx, Zioutas:2020cul, DelleRose:2020pbh, Chao:2020yro, An:2020tcg, Hryczuk:2020jhi, Alhazmi:2020fju, Cacciapaglia:2020kbf, Ko:2020gdg, Baek:2020owl, Okada:2020evk, Choi:2020kch,  Davighi:2020vap, He:2020wjs, Davoudiasl:2020ypv, Chiang:2020hgb, Arcadi:2020zni, Choudhury:2020xui, Ema:2020fit, VanDong:2020bkg,1807899, 1807957, 1807895}. For other interpretations and discussions related to this excess, please refer to \cite{Arias-Aragon:2020qtn, Athron:2020maw, Shoemaker:2020kji, Babu:2020ivd, Miranda:2020kwy, Chigusa:2020bgq, Li:2020naa, Croon:2020ehi, Szydagis:2020isq, Gao:2020wfr, Dessert:2020vxy, Ge:2020jfn, Bhattacherjee:2020qmv, Coloma:2020voz, McKeen:2020vpf, Dent:2020jhf, Bloch:2020uzh, Robinson:2020gfu, Lindner:2020kko, Gao:2020wer, Khan:2020vaf, AristizabalSierra:2020edu, Buch:2020mrg, DiLuzio:2020jjp, Bally:2020yid, Boehm:2020ltd}. In the present work, we adopt the idea of inelastic DM in the context of XENON1T excess within the framework of a well motivated gauge extension of the standard model (SM).

One popular extension of the SM is the implementation of an Abelian gauge symmetry $L_{\alpha}-L_{\beta}$ where $L_{\alpha}$ is the lepton number of generation $\alpha = e, \mu, \tau$. Interestingly, such a gauge extension is anomaly free and can have very interesting phenomenology related to neutrino mass, DM as well as flavour anomalies like the muon anomalous magnetic moment $(g-2)_{\mu}$ \cite{Tanabashi:2018oca}. While there can be three different combination for this gauge symmetry, we particularly focus on $L_{\mu}-L_{\tau}$ gauge symmetry. For earlier works in different contexts, please see \cite{He:1990pn, Baek:2001kca, Ma:2001md, Patra:2016shz} and references therein. Apart from the SM fermions and three right handed neutrinos required for generating light neutrino masses through type I seesaw mechanism \cite{Minkowski:1977sc, GellMann:1980vs, Mohapatra:1979ia}, we have a Dirac fermion which is naturally stable due to the chosen quantum number under the new gauge symmetry. The scalar singlets which break the new gauge symmetry spontaneously also gives masses to the right handed neutrinos. While DM fermion has a bare mass term, one of the scalar singlets give a Majorana mass term splitting the Dirac fermion into two pseudo-Dirac mass eigenstates. These states can be inelastic DM~\cite{TuckerSmith:2001hy, Cui:2009xq, Arina:2012aj, Arina:2012fb, Arina:2011cu} of the universe. If the mass splitting between these two mass eigenstates is appropriately tuned, the heavier component can be long-lived and can comprise a significant fraction of total DM density in the present universe. Here we show that inelastic fermion DM can give rise to the required electron recoil events observed by XENON1T while at the same time being consistent with relic abundance, muon $(g-2)$ and light neutrino mass criteria. Similar idea of addressing muon $(g-2)$ and XENON1T excess within a scalar extension of the SM was recently proposed in \cite{Choudhury:2020xui}. Another recent work, particularly in the context of $L_{\mu}-L_{\tau}$ gauge symmetry, showed that solar neutrinos with such new gauge interactions can not be responsible for XENON1T excess \cite{Amaral:2020tga}. Our proposal in this work provides an alternative way to address the excess in gauged $L_{\mu}-L_{\tau}$ model augmented by inelastic fermion DM.

While inelastic DM as an origin of XENON1T excess has already been pointed out in hidden sector DM models 
(where DM is a SM singlet but charged under hidden sector symmetry and hence interact with the SM particles via kinetic mixing), our framework 
provides a scenario where both DM and SM are charged under the additional gauge symmetry and hence DM-SM interaction happens without kinetic mixing. 
This is a crucial difference which not only leads to a better prospects of detection, specially in the context of several muonic probes to be discussed 
later, but also gives a different DM phenomenology in the context of early universe compared to hidden sector models. Although neutrino mass remains 
disconnected from the numerical analysis of DM related observables, the model incorporates light neutrino masses naturally. Additionally, the model 
also provides a solution to the longstanding muon $(g-2)$ anomaly as mentioned earlier. Combining with all these bounds and requirements, it leaves only a tiny parameter space in our model, making it falsifiable with near future data while the hidden sector DM model, see for example~\cite{Harigaya:2020ckz}, leaves 
a much wider parameter space. \\

\noindent
{\bf Gauged $L_{\mu}-L_{\tau}$ Symmetry}: As mentioned before, we consider an $L_{\mu}-L_{\tau}$ gauge extension of the SM. The SM fermion content with their gauge charges under $SU(3)_c \times SU(2)_L \times U(1)_Y \times U(1)_{L_{\mu}-L_{\tau}}$ gauge symmetry are denoted as follows.

$$ q_L=\begin{pmatrix}u_{L}\\
d_{L}\end{pmatrix} \sim (3, 2, \frac{1}{6}, 0), \; u_R (d_R) \sim (3, 1, \frac{2}{3} (-\frac{1}{3}), 0)$$
$$L_e=\begin{pmatrix}\nu_{e}\\
e_{L}\end{pmatrix} \sim (1, 2, -\frac{1}{2}, 0), \; e_R \sim (1, 1, -1, 0) $$
$$L_{\mu}=\begin{pmatrix}\nu_{\mu}\\
\mu_{L}\end{pmatrix} \sim (1, 2, -\frac{1}{2}, 1), \; \mu_R \sim (1, 1, -1, 1) $$
$$L_{\tau}=\begin{pmatrix}\nu_{\tau}\\
\tau_{L}\end{pmatrix} \sim (1, 2, -\frac{1}{2}, -1), \;  \tau_R \sim (1, 1, -1, -1)$$

Note that the chiral fermion content of the model mentioned above keeps the model free from triangle anomalies. The DM field is represented by a Dirac fermion $\chi_{L,R} \sim (1, 1, 0, \frac{1}{2})$ where the choice of $L_{\mu}-L_{\tau}$ charge is made in such a way that stabilise it without requiring any additional symmetries. In order to break the gauge symmetry spontaneously as well as to generate the desired fermion mass spectrum, the scalar fields are chosen as follows.
$$H=\begin{pmatrix}H^+\\
H^0\end{pmatrix} \sim (1,2,\frac{1}{2},0), \; \phi_1 (\phi_2) \sim (1, 1, 0, 1 (2))$$
While the neutral component of the Higgs doublet $H$ breaks the electroweak gauge symmetry, the singlets break $L_{\mu}-L_{\tau}$ gauge symmetry after acquiring non-zero vacuum expectation values (VEV). Denoting the VEVs of singlets $\phi_{1,2}$ as $v_{1,2}$, the new gauge boson mass can be found to be $M_{Z'}=g_x \sqrt{(v^2_1+4v^2_2)}$ with $g_x$ being the $L_{\mu}-L_{\tau}$ gauge coupling. Please note that, in principle, the symmetry of the model allows a kinetic mixing term between $U(1)_Y$ of SM and $U(1)_{L_{\mu}-L_{\tau}}$ of the form $\frac{\epsilon}{2} B^{\alpha \beta} Y_{\alpha \beta}$ where $B^{\alpha\beta}=  \partial^{\alpha}X^{\beta}-\partial^{\beta}X^{\alpha}, Y_{\alpha \beta}$ are the field strength tensors of $U(1)_{L_{\mu}-L_{\tau}}, U(1)_Y$ respectively and $\epsilon$ is the mixing parameter. This kinetic mixing plays a crucial role in giving rise to the XENON1T excess as we discuss later.

The relevant part of the DM Lagrangian is 
\begin{align}
-\mathcal{L}_Y &= M_{\chi} (\bar{\chi_L} \chi_R + \bar{\chi_R} \chi_L) +  \frac{1}{2}(f_1 \overline{\chi^c_L} \chi_L \phi^*_1+ \nonumber \\
& +f_2 \overline{\chi^c_R} \chi_R \phi^*_1 +{\rm h.c.})
\end{align}
The DM field is identified as $\chi$ which has a bare mass as well as coupling to $\phi_1$. While the bare mass term is of Dirac type, the coupling to $\phi_1$ introduces a Majorana mass term after $\phi_1$ acquires a non-zero VEV. Thus, the Dirac fermion $\chi$ is split into two Majorana fermions $\chi_1, \chi_2$. The DM Lagrangian in this physical basis is 
\begin{widetext}
\begin{align}
\mathcal{L}_{\rm DM} & = \frac{1}{2} \bar{\chi_1} i \gamma^{\mu} \partial_{\mu} \chi_1 -\frac{1}{2} M_1 \bar{\chi^c_1} \chi_1+ \frac{1}{2} \bar{\chi_2} i \gamma^{\mu} \partial_{\mu} \chi_2-\frac{1}{2} M_2 \bar{\chi^c_2} \chi_2 +(i\frac{1}{2}g_x \bar{\chi_2} \gamma^{\mu} \chi_1 Z'_{\mu} +{\rm h.c.})  \nonumber \\
& +\frac{1}{4} g_x \frac{m_-}{M_{\chi}} (\bar{\chi_2} \gamma^{\mu} \gamma^5 \chi_2-\bar{\chi_1} \gamma^{\mu} \gamma^5 \chi_1)Z'_{\mu} +\frac{1}{2}(f_1 \cos^2{\theta}-f_2 \sin^2{\theta}) \bar{\chi_1} \chi_1 \phi_1+\frac{1}{2}(f_2 \cos^2{\theta}-f_1 \sin^2{\theta}) \bar{\chi_2} \chi_2 \phi_1
\end{align}
\end{widetext}
where $M_1 = M_{\chi}-m_+, M_2=M_{\chi}+m_+, m_{\pm} = (m_L \pm m_R)/2, m_{L,R} = f_{1,2} v_{1}$. As will be discussed below, the mass splitting between $\chi_1, \chi_2$ is chosen to be very small $\delta = M_2-M_1 =2m_+ \sim \mathcal{O}(\rm keV)$ in order to give the required fit to XENON1T excess. This ensures $M_1 \approx M_2 \approx M_{\chi}$ while leaving $m_{-}$ as a free parameter. In the above Lagrangian for DM, $\theta$ is a mixing angle given by $\tan{\theta} \approx m_{-}/M_{\chi}$. \\

\noindent
{\bf Light Neutrino Masses}: In order to account for tiny non-zero neutrino masses for light neutrinos, we extend the minimal gauged $U(1)_{L_\mu - L_\tau}$ model with additional neutral fermions as 
$$N_e \sim (1,1,0,0), N_\mu (N_\tau) \sim (1,1,0, 1(-1))$$ 
where the quantum numbers in the parentheses are the gauge charges under $SU(3)_c \times SU(2)_L \times U(1)_Y \times U(1)_{L_{\mu}-L_{\tau}}$ symmetry. Also, the chosen gauge charges of right handed neutrinos do not introduce any new contribution to triangle anomalies. The relevant Yukawa interaction terms are given by
\begin{align}
\mathcal{L} \supset &-\frac{1}{2} M_{ee} \overline{N^c_e} N_e -\frac{1}{2} M_{\mu \tau} \overline{N^c_\mu} N_\tau-
( \lambda_{e\mu} \phi_1^\star \overline{N^c_e} N_\mu + {\rm h.c.} ) \nonumber \\
         & - ( \lambda_{e\tau} \phi_1 \overline{N^c_e} N_\tau + {\rm h.c.} ) -( \lambda_{\mu\mu} \phi_2^\star \overline{N^c_\mu} N_\mu +{\rm h.c.} ) \nonumber \\
           & - ( \lambda_{\tau\tau} \phi_2 \overline{N^c_\tau}
           N_\tau + {\rm h.c.} ) \nonumber \\
           &-\left( Y_{ee} \overline{L_e} \tilde{H} N_e + Y_{\mu \mu} \overline{L_\mu} \tilde{H} N_\mu
             + Y_{\tau \tau} \overline{L_\tau} \tilde{H} N_\tau+ {\rm h.c.} \right) \nonumber \\
           & -\left( Y_{e} \overline{L_e} H e_R + Y_{\mu} \overline{L_\mu} H \mu_R
             + Y_{\tau} \overline{L_\tau} H \tau_R + {\rm h.c.} \right) \nonumber \\
          =& -\frac{1}{2} N^T_\alpha \mathcal{C}^{-1} {M_R}_{\alpha \beta} N_\beta
             - {M_D}_{\alpha \beta} \overline{\nu_\alpha} N_\beta - M_\ell \overline{\ell_L} \ell_R +{\rm h.c.}
\end{align}
Which clearly predict diagonal charged lepton and Dirac neutrino mass matrices $M_\ell, M_D$. Thus, the non-trivial neutrino mixing will arise from the structure of right handed neutrino mass matrix $M_R$ only which is generated by the chosen scalar singlet fields. The ight handed neutrino, Dirac neutrino and charged lepton mass matrices are given by
\begin{align}
M_R =\begin{pmatrix}
               M_{ee}      &  \lambda_{e\mu} v_1
    & \lambda_{e\tau} v_ 1 \\
               \lambda_{e\mu} v_ 1      &  \lambda_{\mu \mu}
v_2     & M_{\mu \tau}  \\
               \lambda_{e\tau} v_1     &  M_{\mu \tau}    &
               \lambda_{\tau \tau} v_2  
               \end{pmatrix}\, \nonumber \\
M_D =\begin{pmatrix}
               Y_{ee} v      &  0    & 0  \\
               0     &  Y_{\mu \mu}v    & 0  \\
               0     &  0    & Y_{\tau \tau} v  
               \end{pmatrix},\,              M_\ell= \begin{pmatrix}
Y_{e} v & 0 & 0\\
0 & Y_{\mu}v & 0 \\
0 & 0 & Y_{\tau} v
\end{pmatrix}
\end{align}
Using type I seesaw approximation $(M_D \ll M_R)$, the light neutrino mass matrix can be found from the following seesaw formula
\begin{align}
m_\nu \simeq - M_D M^{-1}_R M^T_D \,.
\end{align}
Since the charged lepton mass matrix is diagonal in our model, the light neutrino mass matrix can be diagonalised by using the Pontecorvo-Maki-Nakagawa-Sakata (PMNS) mixing matrix as
$$m^{\rm diag.}_\nu = U_{\rm PMNS}^\dagger m_\nu U_{\rm PMNS}^* = \mbox{diag}\{m_1, m_2, m_2 \}$$
where $m_i$ are the light neutrino mass eigenvalues. Since $M_R$ has a very general structure, one can fit the model parameters with light neutrino data in several ways, independently of rest of our analysis. \\

\noindent
{\bf Anomalous Muon Magnetic Moment}: The magnetic moment of muon is given by
\begin{equation}\label{anomaly}
\overrightarrow{\mu_\mu}= g_\mu \left (\frac{q}{2m} \right)
\overrightarrow{S}\,,
\end{equation}
where $g_\mu$ is the gyromagnetic ratio and its value is $2$ for a
structureless, spin $\frac{1}{2}$ particle of mass $m$ and charge
$q$. Any radiative correction, which couples the muon spin to the
virtual fields, contributes to its magnetic moment and is given by
\begin{equation}
a_\mu=\frac{1}{2} ( g_\mu - 2)
\end{equation}
The anomalous muon magnetic moment has been measured very precisely while it has also been predicted in the SM to a great accuracy. At present the difference between the predicted and the measured value is given by 
\begin{equation}
\Delta a_{\mu} = a_{\mu}^{\rm exp} - a_{\mu}^{\rm SM} = (26.1 \pm 7.9)\times 10^{-10}, 
\end{equation}
which shows there is still room for NP beyond the SM (for details see~\cite{Tanabashi:2018oca}). In a recent article, the status of the SM calculation of muon magnetic moment has been updated \cite{Aoyama:2020ynm}. According to this study $\Delta a_{\mu}  = (27.9 \pm 7.6)\times 10^{-10}$ which is a 3.7$\sigma$ discrepancy. Quoting the errors in $3\sigma$ range, we have 
$$\Delta a_{\mu}  = (27.9 \pm 22.8)\times 10^{-10}$$
In our model, the additional contribution to muon magnetic moment comes from one loop diagram mediated by $Z'$ boson. The contribution is given by \cite{Brodsky:1967sr, Baek:2008nz}
\begin{equation}
\Delta a_{\mu} = \frac{\alpha'}{2\pi} \int^1_0 dx \frac{2m^2_{\mu} x^2 (1-x)}{x^2 m^2_{\mu}+(1-x)M^2_{Z'}} \approx \frac{\alpha'}{2\pi} \frac{2m^2_{\mu}}{3M^2_{Z'}}
\end{equation}
where $\alpha'=g^2_x/(4\pi)$. \\

\noindent
{\bf Relic Abundance of DM}: Relic abundance of two component DM in our model $\chi_{1,2}$ can be found by numerically solving the corresponding Boltzmann equations. Let $n_2 = n_{\chi_2} $ and
$n_1=n_{\chi_1}$ are the total
number densities of two dark matter
candidates respectively. The two coupled Boltzmann
equations in terms of $n_2$ and $n_1$ are given below,   
\begin{widetext}
\begin{eqnarray}
\frac{dn_{2}}{dt} + 3n_{2} H &=& 
-\langle{\sigma {\rm{v}}}_{\chi_2 \bar{\chi_2} \rightarrow {X \bar{X}}}\rangle 
\left(n_{2}^2 -(n_{2}^{\rm eq})^2\right)
- {\langle{\sigma {\rm{v}}}_{\chi_2 \bar{\chi_2}
\rightarrow \chi_1 \bar{\chi_1}}\rangle} \bigg(n_{2}^2 - 
\frac{(n_{2}^{\rm eq})^2}{(n_{1}^{\rm eq})^2}n_{1}^2\bigg) -  \langle{\sigma {\rm{v}}}_{\chi_2 \bar{\chi_1} \rightarrow {X \bar{X}}}\rangle 
\left(n_{1} n_2 -n_{1}^{\rm eq} n_{2}^{\rm eq}  \right)\,, \nonumber 
%
\label{boltz-eq1} \\
\frac{dn_{1}}{dt} + 3n_{1} H &=& -\langle{\sigma {\rm{v}}}
_{\chi_1 \bar{\chi_1} \rightarrow {X \bar{X}}}\rangle \left(n_{1}^2 -
(n_{1}^{\rm eq})^2\right) 
+ {\langle{\sigma {\rm{v}}}_{\chi_2 \bar{\chi_2} \rightarrow {\chi_1} \bar{\chi_1}}\rangle} 
\bigg(n_{2}^2 - \frac{(n_{2}^{\rm eq})^2}{(n_{1}^{\rm eq})^2}
n_{1}^2\bigg) - \langle{\sigma {\rm{v}}}_{\chi_2 \bar{\chi_1} \rightarrow {X \bar{X}}}\rangle 
\left(n_{1} n_2 -n_{1}^{\rm eq} n_{2}^{\rm eq}  \right) \,,\nonumber \\
\label{boltz-eq2} 
\end{eqnarray}
\end{widetext}
where, $n^{\rm eq}_i$ is the equilibrium number density
of dark matter species $i$ and $H$ denotes the Hubble parameter.  The thermally averaged annihilation and coannihilation processes ($\chi_i \bar{\chi_i} \rightarrow X \bar{X}$) are denoted by $\langle{\sigma {\rm{v}}} \rangle$ where X denotes all particles to which DM can annihilate into. Since we consider GeV scale DM, the only annihilations into light SM fermions can occur. We consider all the singlet scalars to be heavier than DM masses. Also, the singlet mixing with SM Higgs are assumed to be tiny so that singlet mediated annihilation channels are negligible and only the annihilations mediated by $Z'$ gauge boson dominate. Additionally, the keV scale mass splitting between the two DM candidates lead to efficient coannihilations while keeping their conversions into each other sub-dominant. We have solved these two coupled Boltzmann equations using \texttt{micrOMEGAs} \cite{Belanger:2014vza}. Due to tiny mass splitting, almost identical annihilation channels and sub-dominant conversion processes, we find almost identical relic abundance of two DM candidates. Thus each of them constitutes approximately half of total DM relic abundance in the universe. We constrain the model parameters by comparing with Planck 2018 limit on total DM abundance $\Omega_{\text{DM}} h^2 = 0.120\pm 0.001$ \cite{Aghanim:2018eyx}. Here $\Omega_{\rm DM}$ is the density parameter of DM and $h = \text{Hubble Parameter}/(100 \;\text{km} ~\text{s}^{-1} \text{Mpc}^{-1})$ is a dimensionless parameter of order one.

Although relative abundance of the two DM candidates $\chi_1$ and $\chi_2$ are expected to be approximately the half of total DM relic 
abundance from the above analysis based on chemical decoupling of DM from the SM bath, there can be internal conversion happening between the two DM candidates via processes like $\chi_2 \chi_2 \rightarrow \chi_1 \chi_1, \chi_2 e \rightarrow \chi_1 e$ until later epochs. While such processes keep the total DM density conserved, they can certainly change the relative proportion of two DM densities. It was pointed out by \cite{Harigaya:2020ckz, Baryakhtar:2020rwy} as well as several earlier works including \cite{Finkbeiner:2007kk, Batell:2009vb}. In these works, DM is part of a hidden sector comprising a gauged $U(1)_X$ which couples to the SM particles only via kinetic mixing of $U(1)_X$ and $U(1)_Y$, denoted by $\epsilon$. Thus, although the DM-SM interaction is suppressed by $\epsilon^2$ leading to departure from chemical equilibrium at early epochs, the internal DM conversions like $\chi_2 \chi_2 \rightarrow \chi_1 \chi_1$ can happen purely via $U(1)_X$ interactions and can be operative even at temperatures lower than chemical freeze-out temperature. However, one crucial difference between such hidden sector DM models and our model is that both DM and SM are charged under the new gauge symmetry $U(1)_{L_{\mu}-L_{\tau}}$. And hence both DM-SM interactions as well $\chi_2 \chi_2 \rightarrow \chi_1 \chi_1$ freeze out at same epochs. On the other hand, the interaction $\chi_2 e \rightarrow \chi_1 e$ is suppressed due to kinetic mixing involved and hence it is unlikely to be effective after the above two processes freeze out.

\begin{figure}[h!]
\centering
\includegraphics[scale=0.50]{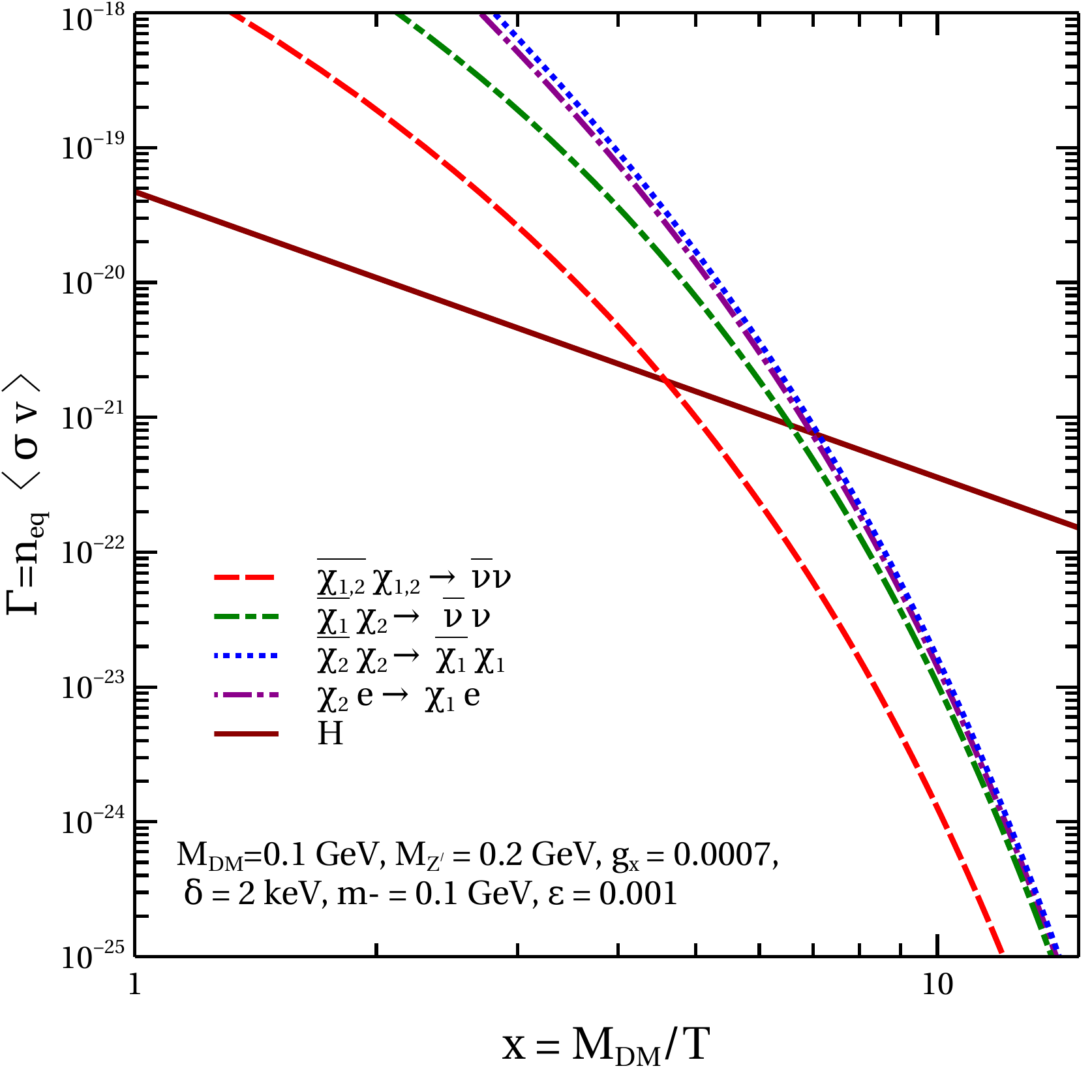}
\caption{Rates of different DM scattering processes in comparison to the Hubble expansion rate.}
\label{Fig0}
\end{figure}

For a quantitative comparison, we estimate the cross sections of different processes relevant for DM mass below 100 MeV as 
\begin{widetext}
\begin{eqnarray}
\sigma (\chi_{1,2} \chi_{1,2} \rightarrow \nu \bar{\nu}) & \ = \ &\frac{g^4_x m^2_-s}{96 \pi M^2_{1,2} (s-M^2_{Z'})^2} \sqrt{1-\frac{4 M^2_{1,2}}{s}} \nonumber \\
\sigma (\chi_{1} \chi_{2} \rightarrow \nu \bar{\nu}) & \ = \ & \frac{g^4_x \left ( 2s + (M_1+M_2)^2 \right)}{48 \pi \left ( s - (M_1+M_2)^2 \right) (s-M^2_{Z'})^2} \sqrt{ M^4_1 + (s-M^2_2)^2 -2M^2_1 (s+M^2_2)} \nonumber \\
\sigma (\chi_2 \chi_2 \rightarrow \chi_1 \chi_1) & \ = \ & \frac{g^4_x}{1536 \pi M^4_{Z'} s (s-4M^2_2)} f_1 (M_1, M_2, s, M_{Z'}, m_-) \nonumber \\
\sigma (\chi_2 e \rightarrow \chi_1 e) & \ = \ & \frac{g^2_x g^2 \epsilon^2}{8\pi \left ( (s-M^2_2-m^2_e)^2-4m^2_e M^2_2 \right)} f_2 (M_1, M_2, s, M_{Z'})
\end{eqnarray}
\end{widetext}
where $f_1, f_2$ are functions of model parameters, the details of which are skipped here for simplicity, but taken into account in the numerical calculations. It is important to note that the first three processes depend upon gauge coupling $g_x$ in the same fashion while the last one depends on $\epsilon$ as well. Clearly, for our chosen values of $g_x, \epsilon$ we have similar $g^4_x$ and $g^2_x g^2 \epsilon^2$ where $g$ is the electroweak gauge coupling. For a comparison, we show the rates of these processes in comparison to Hubble expansion rate in figure \ref{Fig0}. We have used $g_x = 0.0007, \epsilon=0.001, M_1 = 0.1$ GeV, $ \delta = 2$ keV, $M_{Z'} = 0.2$ GeV. Clearly, the internal DM conversion processes decouple almost simultaneously with the DM annihilation and coannihilation processes, as expected. Therefore, the estimate of DM abundance based on the chemical decoupling is justified in our setup.

Since the mass splitting between $\chi_2$ and $\chi_1$ is kept at keV scale $\delta \sim \mathcal{O}(\rm keV)$, there can be decay modes like $\chi_2 \rightarrow \chi_1 \nu \bar{\nu}$ primarily mediated by $Z'$. If both the DM components are to be there in the present universe, this lifetime has to be more than the age of the universe that is $\tau_{\chi_2} >\tau_{\rm age} \approx 4 \times 10^{17}$ s. The decay width of this process is $\Gamma_{\chi_2 \rightarrow \chi_1 \nu \bar{\nu}} \approx g^4_x \delta^5/(160 \pi^3 M^4_{Z'})$. Thus, imposing the lifetime constraint on heavier DM component puts additional constraints on the model parameters. \\

\noindent
{\bf XENON1T Excess}: Similar to the proposal in \cite{Harigaya:2020ckz}, here also we consider the down-scattering of heavier DM component $\chi_2 e \rightarrow \chi_1 e$ as the process responsible for XENON1T excess of electron recoil events near 1-3 keV energy \cite{Aprile:2020tmw}.

For a fixed DM velocity $v$,the differential cross section is given by
 \begin{equation}
 \frac{d\sigma v}{d E_r} = \frac{\sigma_e}{2 m_e v} \int_{q-}^{q+} a^2_0 q dq|F(q)|^2 K(E_r,q)
 \label{Event}
 \end{equation}	
where $m_e$ is the electron mass, $a_0 = \frac{1}{\alpha m_e}$ is the Bohr radius, $\alpha = \frac{e^2}{4 \pi}=\frac{1}{137}$ is the fine structure constant, $E_r$ is the recoil energy, $q$ is the transferred momentum, $K(E_r, q)$ is the atomic excitation factor, and $\sigma_e$ is the free electron cross section. The atomic excitation factor is taken from \cite{Roberts:2019chv}. We assume the DM form factor to be unity. The free electron cross-section is given by
\begin{equation}
\sigma_e = \frac{16 \pi \alpha_z \alpha_{x} \epsilon^2 m^2_e  }{M^4_{Z'}}
\end{equation}
where $\alpha_z=\frac{g^2}{4 \pi}$, $\alpha_{x}=\frac{g^2_x}{4 \pi}$ and $\epsilon$ is the kinetic mixing parameter between $Z$ and $Z'$ mentioned earlier which we take to be $\epsilon \leq 10^{-3}$. Here in the inelastic scattering case, the limits of integration in Eq.~\eqref{Event} are determined depending on the relative values of recoil energy ($E_r$) and the mass splitting between the DM particles ($\delta=M_2-M_1$). It should be noted that $\sigma_e$ is independent of DM mass as the reduced mass of DM-electron is almost equal to electron mass for GeV scale DM mass we are considering.
	
For $E_r \geq \delta$
\begin{equation}
q_\pm=M_{2} v \pm \sqrt{M^2_{2} v^2 -2M_{2}(E_r-\delta)}
\end{equation}
And for $E_r \leq \delta$
\begin{equation}
q_\pm=\sqrt{M^2_{2} v^2 -2M_{2}(E_r-\delta)} \pm M_{2} v 	
\end{equation}
The differential event rate for the inelastic DM scattering with electrons in xenon is given by
\begin{equation}
\frac{dR}{dE_r}=n_T n_{\chi_2} \frac{d \sigma v}{d E_r}
\end{equation}
where $n_T=4\times10^{27}$ $ {\rm Ton}^{-1}$ is the number density of xenon atoms and $n_{\chi_2}$ is the density of the dark matter $\chi_2$. As mentioned before $n_{\chi_2} \approx n_{\chi_1} \approx n_{\rm DM}/2$. 

\begin{figure}[h!]
\centering
\includegraphics[scale=0.48]{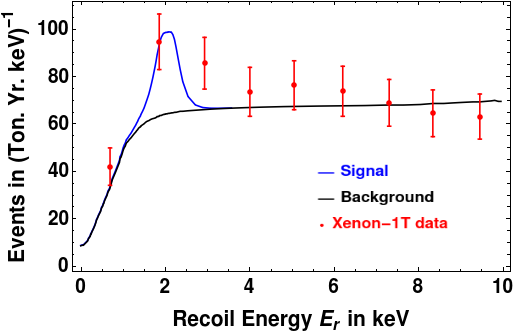}
\caption{Fit to XENON1T data with inelastic fermion DM in our model.}
\label{Fig1}
\end{figure}

\begin{figure*}
\centering
\includegraphics[scale=0.48]{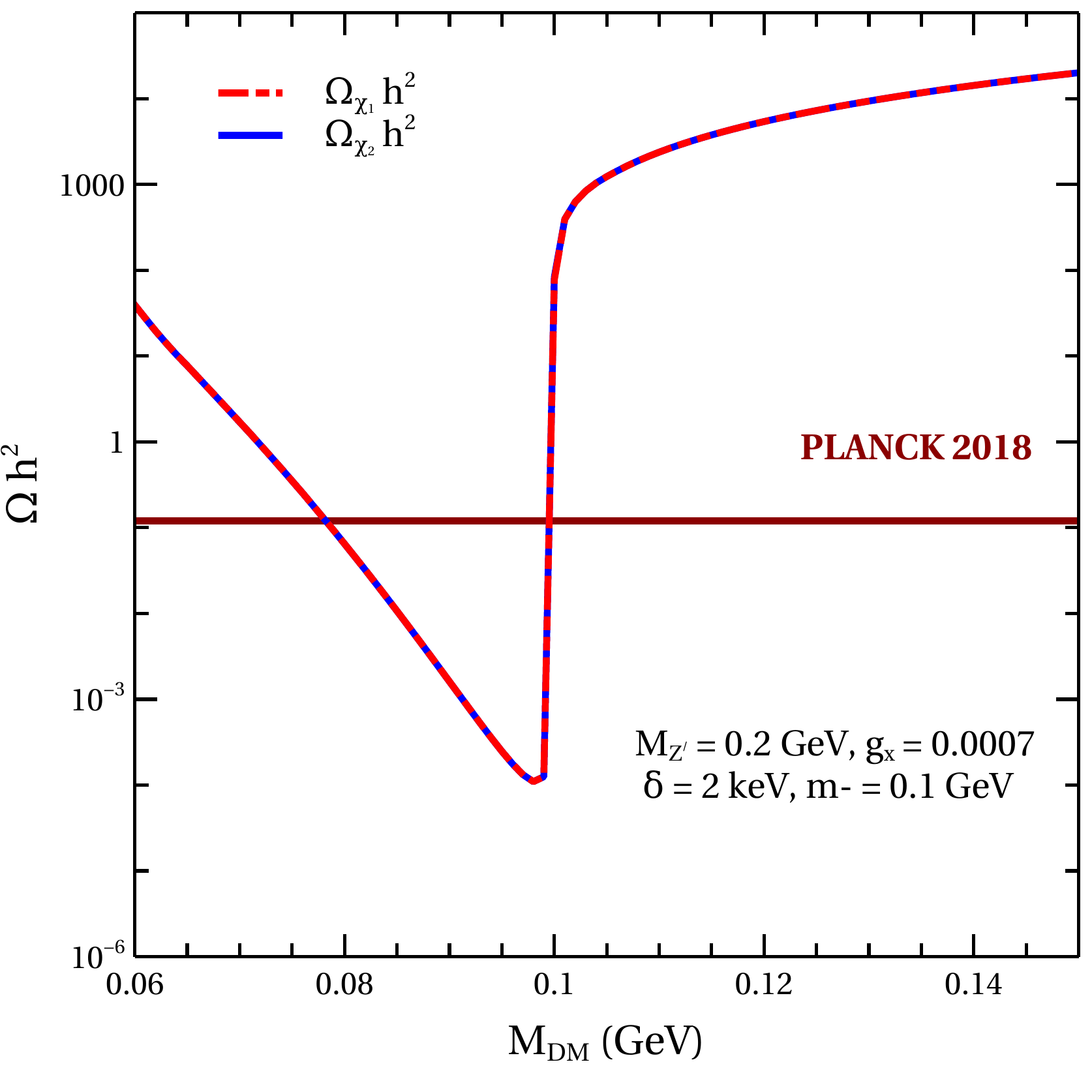}
\includegraphics[scale=0.48]{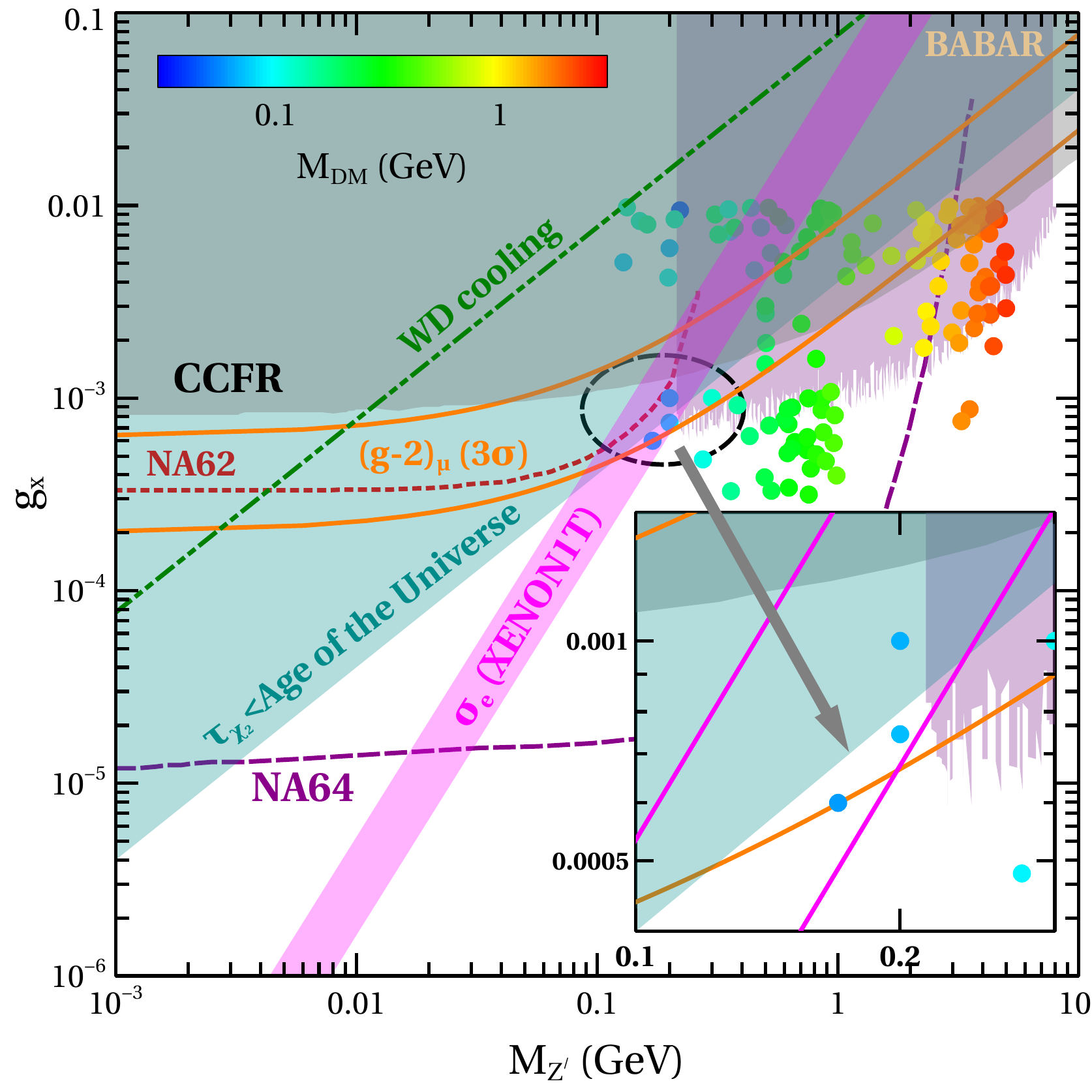}
\caption{Left panel: Variation of relic abundance with DM mass for fixed benchmark values of relevant model parameters. Right panel: Summary plot showing the final parameter space from all relevant constraints. Here $M_{\rm DM} \sim M_1 \sim M_2$ is the mass of two almost degenerate DM candidates.}
\label{Fig2}
\end{figure*}

\noindent
{\bf Results and Conclusion}: We first fit our model with XENON1T data using the methodology described above. The result is shown in Fig.~\ref{Fig1}. The mass splitting is taken to be $\delta =2$ keV while heavier DM mass is taken to be 0.1 GeV consistent with all relevant constraints. DM velocity is taken to be $v \approx 5 \times 10^{-3}$, consistent with its non-relativistic nature. The other relevant parameters used in this fit are $g_x = 7\times10^{-4}, M_{Z'} = 0.2$ GeV, $\epsilon = 10^{-3}$ which corresponds cross section $\sigma_e = 4.8 \times 10^{-17} \; {\rm GeV}^{-2}$. As we discuss below, this choice of parameters is also consistent with all other relevant bounds.

We then calculate the relic abundance of two DM candidates $\chi_2, \chi_1$ using the procedures mentioned above. The left panel of Fig.~\ref{Fig2} shows the variation of DM relic abundance with DM mass for a set of fixed benchmark parameters. Clearly, due to tiny mass splitting between two DM candidates and identical gauge interactions, their relic abundances are almost identical. The DM annihilation due to s-channel mediation of $Z'$ gauge boson is clearly visible from this figure where correct relic of DM is satisfied near the resonance region $M_{\rm DM} \approx M_{Z'}/2$.

Final result is summarised in the right panel plot of Fig.~\ref{Fig2} in terms of parameter space $g_x-M_{Z'}$. The parameter space satisfying anomalous muon magnetic moment in $3\sigma$ is shown within the orange coloured solid lines. The grey shaded region corresponds to the parameter space excluded by upper bound on cross sections for $\nu N \rightarrow \nu N \mu \bar{\mu}$ measured by CCFR \cite{Altmannshofer:2014pba}. This constraint on $g_x-M_{Z'}$ plane arises purely due to the fact that $L_{\mu}-L_{\tau}$ gauge boson can contribute to this neutrino trident process. It completely rules out the parameter space satisfying $(g-2)_{\mu}$ at $3\sigma$ beyond $M_{Z'} \gtrsim 1$ GeV. The shaded region of light green colour shows the parameter space where the bound on lifetime of heavier DM $\chi_2$ mentioned earlier is not satisfied and hence ruled out. Clearly, this lifetime bound is stronger than the CCFR bound for $M_{Z'} \lesssim 1$ GeV. The pink solid band corresponds to $\sigma_e =10^{-17}-10^{-16} \; {\rm GeV}^{-2}$ required to fit the XENON1T excess for the chosen DM velocity $\mathcal{O}(10^{-3})$ and DM mass around 0.1 GeV. The strongest bound in the high mass regime of $Z'$ comes from BABAR observations for $4\mu$ final states \cite{TheBABAR:2016rlg}, as shown by the light pink shaded region in right panel plot of Fig.~\ref{Fig2}. Interestingly, all these bounds allow a tiny part of the parameter space near $M_{Z'} \approx 0.1$ GeV (see inset of right panel plot in Fig.~\ref{Fig2}). While future experiments like NA62 at CERN just falls short of being sensitive to this tiny region \cite{Krnjaic:2019rsv} (red dashed line in right panel plot of Fig.~\ref{Fig2}), the NA64 experiment at CERN is sensitive to the entire parameter space favoured from DM requirements (dashed line of magenta colour in right panel plot of Fig.~\ref{Fig2}) \cite{Gninenko:2014pea, Gninenko:2018tlp}. Similar to NA64, the future ${\rm M}^3$ experiment at Fermilab is also sensitive to most part of our parameter space \cite{Kahn:2018cqs} though we do not show the corresponding sensitivity curve in our plot here.

We also use the strong astrophysical bounds from white dwarf (WD) cooling on such light gauge bosons \cite{Bauer:2018onh}. This arises as the plasmon inside the WD star can decay into neutrinos through off-shell $Z'$ leading to increased cooling efficiency. This leads to a bound in the $g_x-M_{Z'}$ parameter space as \cite{Kamada:2018zxi}
$$ \left ( \frac{g_x}{7.7 \times 10^{-4}} \right)^2 \left( \frac{10 \; {\rm MeV}}{M_{Z'}} \right)^2 \lesssim 1$$
However, in the region of our interest (triangular region allowed from CCFR and lifetime bounds), the WD cooling constraint remains weaker compared to other relevant bounds, as can be seen from the green dotted line in Fig.~\ref{Fig2} (right panel).

We then consider the cosmological bounds on such light DM and corresponding light mediator gauge boson $Z'$. A light gauge boson can decay into SM leptons at late epochs (compared to neutrino decoupling temperature $T^{\nu}_{\rm dec} \sim \mathcal{O}(\rm MeV)$ increasing the effective relativistic degrees of freedom which is tightly constrained by Planck 2018 data as ${\rm N_{eff}= 2.99^{+0.34}_{-0.33}}$ \cite{Aghanim:2018eyx}. As pointed out by the authors of \cite{Kamada:2018zxi, Ibe:2019gpv, Escudero:2019gzq}, such constraints can be satisfied if $M_{Z'} \gtrsim \mathcal{O}(10 \; {\rm MeV})$. As can be seen from the right panel plot in Fig.~\ref{Fig2}, the lifetime requirement of $\chi_2$ already puts a much stronger bound in the region of our interest. Similar bound also exists for thermal DM masses in this regime which can annihilate into leptons. As shown by the authors of \cite{Sabti:2019mhn}, such constraints from the big bang nucleosynthesis (BBN) as well as the cosmic microwave background (CMB) measurements can be satisfied if $M_{\rm DM} \gtrsim \mathcal{O}(1 \; {\rm MeV})$. On the other hand, constraints from CMB measurements disfavour such light sub-GeV thermal DM production in the early universe through s-channel annihilations into SM fermions \cite{Aghanim:2018eyx}. As shown by the author of \cite{Foldenauer:2018zrz} in the context of $U(1)_{L_{\mu}-L_{\tau}}$ gauge model with sub-GeV DM, such CMB bounds can be satisfied for the near resonance region $M_{Z'} \approx 2M_{\rm DM}$ along with correct relic. Specially, in the scenario with keV mass splitting between two DM candidates, the CMB bound on DM annihilation rate into electrons remains weaker compared to lifetime bound as can be checked by comparing the exclusion plots in \cite{Foldenauer:2018zrz} with the ones shown in our work.

Finally, we perform a random scan for relic abundance of two component DM so that their combined relic satisfy the criteria for observed DM relic abundance. This is shown in terms of scattered points in right panel plot of Fig.~\ref{Fig2} where the colour coding is used to denote DM mass. In this random scan, apart from varying $g_x, M_{Z'}$ we also vary DM mass $M_{\rm DM} \sim M_1 \sim M_2$ in the range $(0.05, 3)$ GeV and the other free parameter $m_{-}$ in the range $(0.1, 1)$ GeV while keeping the tiny mass splitting fixed at $\delta =2$ keV. Clearly, only a very few points fall in the small triangular region allowed from all constraints and requirements. The density of these points inside the triangular region will increase for a bigger scan size. Since only a tiny region of parameter space is allowed in this model, more precise measurements of $(g-2)_{\mu}$ will be able to confirm or rule out this model as its possible explanation. Also, the lifetime bound can be relaxed by choosing exotic $L_{\mu}-L_{\tau}$ charge of DM, allowing more parameter space towards upper part of the currently allowed region, seen from inset of right panel plot in Fig.~\ref{Fig2}. This will also bring the parameter space of our model within the sensitivity of future experiment NA62 at CERN \cite{Krnjaic:2019rsv}. However, such exotic charge will also require additional scalar singlets (in order to split the Dirac fermion DM into two pseudo-Dirac fermions) which do not play any role in neutrino mass generation and hence we do not discuss in the context of this minimal model presented here. Future measurements by XENON1T collaboration as well as other future experiments mentioned above will give a clearer picture on the feasibility of this model. We leave a detailed model building and phenomenological study of such low mass DM scenario in the context of electron recoil signatures as well as different possible origin of light neutrino masses and flavour anomalies to future works.\\

\acknowledgements
DB acknowledges the support from Early Career Research Award from the Department of Science and Technology - Science and Engineering research Board (DST-SERB), Government of India (reference number: ECR/2017/001873). SM thanks Anirban Karan for useful discussions. DN thanks Anirban Biswas for useful discussions.

\end{document}